\documentclass[aps,pra,showpacs,twocolumn,superscriptaddress]{revtex4-1}
\usepackage{graphicx}
\usepackage{dcolumn}
\usepackage{bm}
\usepackage{amssymb}
\usepackage{amsmath}
\usepackage{epsfig}
\usepackage{hyperref}
\usepackage{times}
\usepackage{subfigure}

\begin{document}

\title{Generation of Bose-Einstein Condensates' Ground State Through Machine Learning}
\author{Xiao Liang}
\affiliation{Laboratory of Quantum Information, University of Science and Technology of China, Hefei, 230026, China}
\affiliation{Synergetic Innovation Center of Quantum Information and Quantum Physics, University of Science and Technology of China, Hefei, 230026, China}
\author{Sheng Liu}
\affiliation{Laboratory of Quantum Information, University of Science and Technology of China, Hefei, 230026, China}
\affiliation{Synergetic Innovation Center of Quantum Information and Quantum Physics, University of Science and Technology of China, Hefei, 230026, China}
\author{Yan Li}
\email{yli@phy.ecnu.edu.cn}
\affiliation{Department of Physics, East China Normal University, Shanghai, 200241, China}
\author{Yong-Sheng Zhang}
\email{yshzhang@ustc.edu.cn}
\affiliation{Laboratory of Quantum Information, University of Science and Technology of China, Hefei, 230026, China}
\affiliation{Synergetic Innovation Center of Quantum Information and Quantum Physics, University of Science and Technology of China, Hefei, 230026, China}
\date{\today }

\begin{abstract}
We show that both single-component and two-component Bose-Einstein condensates' (BECs) ground states can be simulated by deep convolutional neural networks of the same structure. We trained the neural network via inputting the coupling strength in the dimensionless Gross-Pitaevskii equation (GPE) and outputting the ground state wave-function. After training, the neural network generates ground states faster than the method of imaginary time evolution, while the relative mean-square-error between predicted states and original states is in the magnitude between $10^{-5}$ and $10^{-4}$. We compared the eigen-energies based on predicted states and original states, it is shown that the neural network can predict eigen-energies in high precisions. Therefore, the BEC ground states, which are continuous wave-functions, can be represented by deep convolution neural networks.
\end{abstract}

\maketitle
\section{Introduction}
Because the analytical solutions of non-linear Hamiltonians are difficult to be found, investigating many-body systems relies heavily on numerical simulations. In the aspects of many-body physics, some methods such as matrix product state (MPS) \cite{1} and density matrix renormalization group (DMRG) \cite{2,3} have shown effectiveness in solving eigen-states of one-dimensional or two-dimensional chain systems \cite{4}. For more than one dimension systems, tensor network states \cite{5,6,7,8,9,10} and quantum Monte Carlo methods \cite{11,12,13,14} are widely used. 

Nowadays artificial intelligence has shown talents in playing GO \cite{15}. In the last decade, machine learning technology has gained more and more interests in solving computational problems \cite{16,17,18,19,20}. Several works have investigated speeding up computation with the help of artificial neural network (ANN), for example using ANN to optimize density-functional theory (DFT) is heavily investigated \cite{21,22,23,24,25}. Recently, Restricted Boltzmann Machine (RBM) is investigated to find the ground state of latticed systems \cite{26}, and the RBM representation ability is further investigated in \cite{27}. Futhermore, the effectiveness of RBM raises interest in comparing neural network representations to traditional quantum state representations \cite{28}.  Besides RBM, more advanced neural network such as convolutional neural network has shown effectiveness on distinguishing phases of many-body systems \cite{29}. 

The difference between ANN and solving Hamiltonian is that ANN accepts inputs and outputs as features and tries to find out mathematical relations between these features without using governing equations. It has been shown that ANN is powerful in pattern recognitions, such as categorizing huge number of images. Training the neural network is optimizing the distance between predicted features and real features. The efficiency of the training process depends on both the optimizing method and whether the structure of the neural network is suitable to ``learn" the features. It has been shown that the wave-functions of  latticed systems such as Ising model and antiferromagnetic Heisenberg model can be represented by RBM. It is naturally raised the question that can neural network represent continuous systems? 

Based on quantum mechanics, the wave-function contains the complete information of a quantum system. The wave-functions are obtained by solving Schrodinger equations, while the dynamics of Bose-Einstein condensates (BECs) \cite{30} is governed by Gross-Pitaevskii equation (GPE) \cite{31,32,33}. Nowadays, one of the methods for numerically solving the ground state of GPE is imaginary time evolution \cite{34}. Since the initial state is evolving on imaginary time step, after many interations only the lowest energy part of the initial wave-function dominates. Here, we speed up the computation process of solving GPEs by training neural networks that generate ground states of single-component and two-component BECs in the cases of both one and two dimensions. A better initial state rather than the Gaussian function can effectively reduce iteration times. Instead of inputting features and outputting classification labels, we train the neural network in the way to input Hamiltonian coefficients and output the ground state wave-functions. 

\section{Neural Network Implementation}
\begin{figure*}
\includegraphics[width=0.95\textwidth]{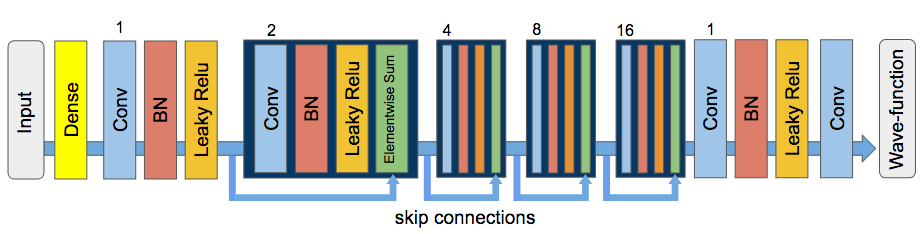}
\caption{ (Color online) Our neural network is made up of several blocks. Each block consists of  a convolutional layer, batch normalization and the Leaky relu activation. The input is the feature that we are interested in the Hamiltonian, such as the coupling coeffcient. The input is transformed into a high dimensional vector by a dense layer, namely, a fully connected layer. The convolutional layer in the first and the last block has the dilation rate of identity,  and the convolutional layers in the intermediate blocks has  the dilation rate of 2, 4, 8 and 16. Dilation rate that is greater than identity is benefit to learn the wave-functions in the larger scale.  The last convolutional layer outputs the wave-function densities, and it has the dilation rate of identity. The output channel of the last layer is either one for single-component BEC or two for two-component BECs.  Since the network has eight layers in total, we use elementwise sum after each block to keep gradients flowing properly. }
\label{fig1}
\end{figure*}
The dynamics of a 2-dimensional BEC is governed by the following GPE,
\begin{equation}
i\hbar\frac{\partial}{\partial t}\psi=\left[ -\frac{\hbar^2}{2m}({\partial}_x^2+{\partial}_y^2)+V(x,y)+g{|\psi|}^2\right]\psi.
\label{1}
\end{equation}

For simplicity, we consider the dimensionless equation that assuming $m=1$ and $\hbar=1$. Under a dimensionless harmonic potential $V(r)=\frac{1}{2}(x^2+y^2)$, the coefficient that matters is only the coupling strength $g$. Since the ground state of GPE is a real function, the weights of the neural network are real numbers. Meanwhile the outputs of the neural network are real distributions.

We set up a deep convolution neural network to learn the ground states of one-dimensional and two-dimensional BECs. A convolution neural network uses filters to scan the feature surface, the relations between adjacent feature sites can be efficiently ``learned" by several filters scanning simultaneously. When the neural network contains tens to hundreds of convolution layers, there has the deep convolution network, such neural network excels at pattern recognition jobs such as image classification, speech recognition and language translation. 

Here we use the deep convolution network in a reversed way to that in classification problems. The structure of the network is depicted in Fig.\ref{fig1}. The first layer is a dense layer, it accepts the inputs of the coupling strength and outputs a high dimensional vector. The output is followed by stacks of convolutional blocks. In the end of the network, a convolution layer outputs a matrix that has the same dimension of the wave-function and the value on each element ranges from 0 to 1, such output forms a distribution in the continuous space.

\section{Results of Single-Component BEC}
We trained the deep convolution network with coupling strengths in the range of $[0,500]$ using 50000 uniformly distributed samples. The samples were generated by the Trotter-Suzuki code \cite{35}. In one-dimensional conditions, there are 512 points in the position space that $x\in[-12,12]$, and each sample is obtained after $10^5$ iterations with a time step of $10^{-4}$. When training the neural network, we randomly select 5000 samples as the validation set, and the remaining 45000 samples were used for training. The distance between the predicted wave-function and the original imaginary time evolved wave-function is the mean-squared-error between two distributions, such distance is calculated by $\int {|\psi(x)_{\text{pred}}-\psi(x)_{\text{original}}|}^2dx$. After training, the distance on either training set or validation set is reduced to the magnitude of $10^{-5}$.

\begin{figure}
\includegraphics[width=0.45\textwidth]{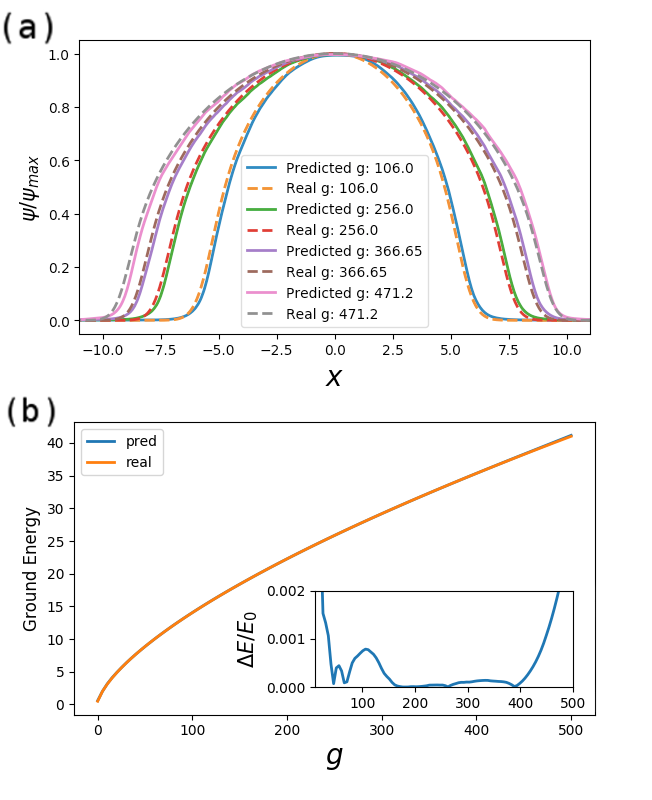}
\caption{ (Color online) (a) The ground state wave-functions generated by the neural network and the imaginary time evolution. Each wave-function is normalized on its maximum value and we keep 512 points in the $x$-space. (b) The comparisons of ground energies generated by neural network and imaginary time evolution. The subfigure depicts the energy error $(E_{\text{pred}}-E_0)/E_0$. }
\label{fig2}
\end{figure}

Our results for one-dimensional BEC is depicted in Fig.\ref{fig2}. In Fig.\ref{fig2}(a) we compare the wave-functions obtained by neural network preditions with imaginary time evolutions. Since the neural network is trained by ground states that $g\in[0,500]$, the neural network predicts ground states in high precisions. To further evaluate the quality of the neural network, we compare ground energies based on the predicted state and the imaginary time evoluted state. The ground energies $E_{0}$ were calculated by
\begin{equation}
E_{0}=\frac{\langle \psi| \hat{H}|\psi\rangle}{\langle\psi|\psi\rangle},
\label{2}
\end{equation}
where the ground energies were evaluated in Fig.\ref{fig2}(b). We use the relative energy error $|E_{\text{predict}}-E_0|/E_0$ to reveal the quality of the predicted states, where $E_{\text{predict}}$ is the ground energy calculated by the predicted state and $E_0$ is the original ground energy. As Fig.\ref{fig2}(b) depicts, the energy error remains in the magnitude of $10^{-3}$ in most of $g$. In the marginal area of the training set that $g=500$, the relative energy error is 0.0034936 and the energy difference is 0.1432696. When $g=0$ the relative energy error is 0.1645099, as the ground energy is 0.5. 

Next we will use the neural network to ``learn" two dimensional states. We also use the Trotter-Suzuki code to generate the training dataset.  50000 samples are prepared in the range of $g\in[0,500]$ uniformly and the position space we are interested in is the squared area that $x,y\in[-7.5,7.5]$. In both $x$-direction and $y$-direction there are 256 points and each sample is obtained after 8000 iterations with the time step of $10^{-3}$. Since the wave-function is two-dimensional, the convolution layers in our neural network are two-dimensional, while the structure of the neural network remains unchanged. The distance to be minimized is then the mean-squared-error calculated on two dimensions. The training process is similar to one-dimensional conditions. After training, the mean-squared-error for either the training set or the validation set is reduced to the magnitude between $10^{-4}$ and $10^{-5}$. We choose neural network after the training epoch that has the minimum validation error.
\begin{figure}
\includegraphics[width=0.4\textwidth]{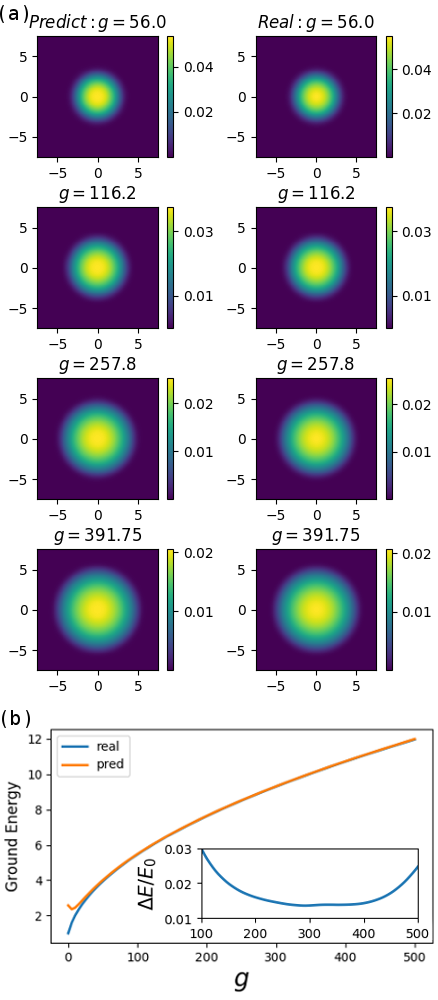}
\caption{ (Color online) (a) The normalized ground state density distribution generated by the neural network (left column) and the imaginary time evolution (right column). We keep 256 points for both $x$ direction and $y$ direction. (b) The comparisons of ground energies generated by neural network and imaginary time evolution. When $g$ is close to zero the predicted energy is far from the original energy. The subfigure depicts the energy error $(E_{\text{pred}}-E_0)/E_0$ where the energy error is in the magnitude of $10^{-2}$. }
\label{fig3}
\end{figure}

Our results of two-dimensional BECs are depicted in Fig.\ref{fig3}. As depicted in Fig.\ref{fig3}(a), the distributions of neural network predited states and imaginary time evoluted states are similar. The comparisons of energies are depicted in Fig.\ref{fig3}(b). When $g$ approaches to zero, the predicted energy is far from the original energy.  When $g$ is close to zero, the spread of wave-function is small compared to our interested area on $x-y$ plane. This makes the training data biased to smaller values, therefore the predicted values are smaller than the original ones. Estimating the eigen-energy using a smaller valued wave-function leads to a higher energy, due to the normalization process. Therefore for two-dimensional states, larger coupling strength leads to wider spread states, which makes the neural network more precisely in predictions.

\section{Results of Two-Component BEC}
\begin{figure*}
\includegraphics[width=0.70\textwidth]{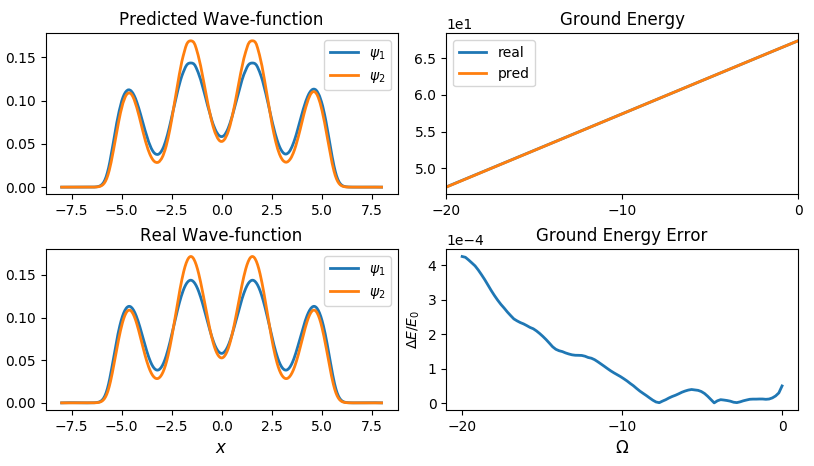}
\caption{ (Color online) When $\Omega=-1$, the neural network predicted states and the real states are depicted on left column. In $x$ direction we keep 512 points. On the right column the energies and the energy errors are depicted in the range of $\Omega\in[-20,0]$. }
\label{fig4}
\end{figure*} 

We continue to investigate that whether neural network can predict two-component BEC states. The ground states of two-component BECs are determined by the coupling strengths of each component ($g_{11}$ and $g_{22}$), the coupling strength between two components ($g_{12}$) and the Rabi coupling coefficient ($\Omega$). The dimensionless GPE of a two-component BEC is
\begin{equation}
i\frac{\partial}{\partial t}\begin{bmatrix}\psi_1\\ \psi_2\end{bmatrix}=\begin{bmatrix}H_1 & \frac{\Omega}{2}\\ \frac{\Omega}{2} & H_2\end{bmatrix}\begin{bmatrix}\psi_1\\ \psi_2\end{bmatrix},
\label{3}
\end{equation}
where $H_1$ and $H_2$ are Hamiltonians of each component that
\begin{equation}
\begin{split}
&H_1=T_1+V_1+g_{11}{|\psi_1|}^2+g_{12}{|\psi_2|}^2,\\
&H_2=T_2+V_2+g_{22}{|\psi_2|}^2+g_{12}{|\psi_1|}^2,
\end{split}
\label{4}
\end{equation}
where $T_{1(2)}$ denotes momentum energy and $V_{1(2)}$ is the potential respectively. To demonstrate the capability of neural network, we investigate the ground states in the range of $\Omega\in[-20,0]$, while $g_{11},g_{12},g_{22}=100(1.03,1,0.97)$. Since there are two components, our neural network has to output two distributions under each input of $\Omega$. 

Firstly we train the neural network using one-dimensional states. The potential is $V(x)=0.5x^2+24\text{cos}^2x$ and our interested area is $x\in[-8,8]$ with 512 points. Since the range of $\Omega\in[-20,0]$ is small, we prepare 13000 samples using the Trotter-Suzuki code, each sample is generated after $10^5$ iterations with a time step of $10^{-4}$. Since the wave-function changes faster as $\Omega$ is close to zero, besides sampling 10000 points uniformly in the range of $\Omega\in[-20,0]$, we additionally sample 3000 points in the range of $\Omega\in[-2,0]$. 1300 samples are randomly picked as the validation set. After training the neural network, the mean-squared-error for both training set and validation set are in the magnitude of $10^{-6}$. In Fig.\ref{fig4}, it is shown that when $\Omega=-1$, the predicted wave-functions are identical to the real wave-functions. To quantify the quality of the predicted wave-function, we compare the ground energies calculated by these wave-functions. The ground energy is calculated as
\begin{equation}
E=\begin{bmatrix}\psi_1 & \psi_2\end{bmatrix}\begin{bmatrix}H_1 &\frac{\Omega}{2}\\ \frac{\Omega}{2} & H_2\end{bmatrix}\begin{bmatrix}\psi_1 \\ \psi_2\end{bmatrix}.
\label{5}
\end{equation}

As depicted in Fig.\ref{fig4}, the error of ground energies is in the magnitude of $10^{-4}$. In the marginal area that $\Omega$ is close to -20, the energy error increases due to the lack of samples. Because of the additional 3000 samples, the energy error remains lower when $\Omega$ is close to zero.

\begin{figure*}
\includegraphics[width=0.70\textwidth]{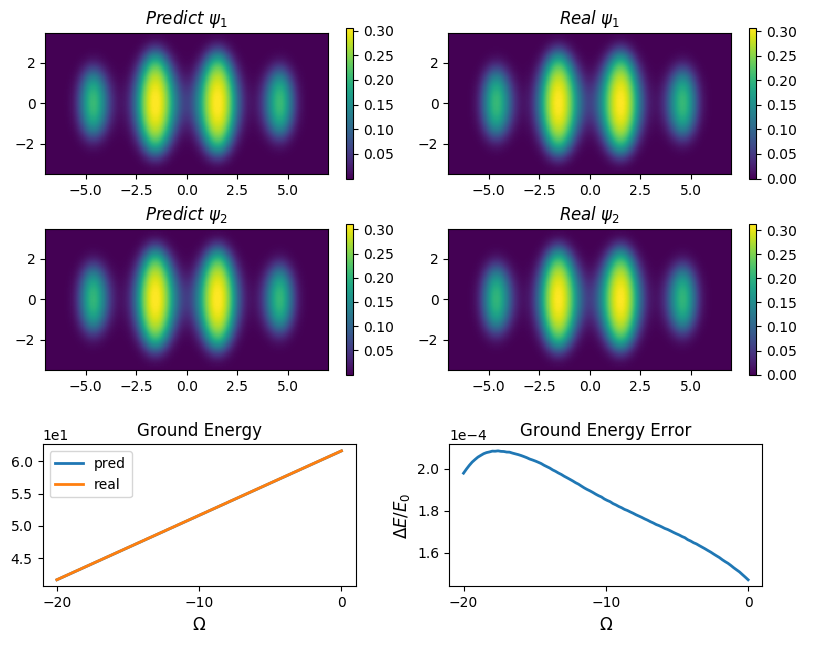}
\caption{ (Color online) For 2-dimensional BECs, when $\Omega=-3.12$, both $\psi_1$ and $\psi_2$ that predicted by neural network are nearly identical to the real ones. We keep 256 points for $x$ direction and 128 points for $y$ direction. The energy error is in the magnitude of $10^{-4}$ under all possible $\Omega$. }
\label{fig5}
\end{figure*}

Fig.\ref{fig5} depicts the two-dimensional conditions. The neural network is also trained by 13000 samples with 1300 samples for validation, each sample is generated by 8000 iterations with a time step of $10^{-3}$. The potential is $V(x,y)=0.5(x^2+5y^2)+\text{cos}^2x$. Since the confinement in $y$-direction is stronger than that in $x$-direction, our interested area is $x\in[-7,7]$ with 256 points and $y\in[-3.5,3.5]$ with 128 points. When $\Omega=-3.12$, as depicted in the figure the predicted states are nearly identical to the real states. The energy error in the range of $\Omega\in[-20,0]$ is in the magnitude of $10^{-4}$. 

Why the energy errors are lower than that of single-component BEC? Because we use 11700 samples in the small range of $\Omega\in[-20,0]$, the dataset is denser than that used for single-component BEC.

\section{Conclusion and Discussion}
We have shown that continuous wave-functions like ground states of BEC can be ``learned" and simulated by deep convolution neural networks. Besides the fact that latticed systems can be simulated by neural networks like RBM, since that convolution network is good at grasping relations between adjacent features, here we show the systems with continuous and smooth distributions can be simulated by convolution neural networks. 

The convolution neural network we trained predicts ground states in high precisions when the inputting coupling strength is in the range of the training set. However when the input feature is far from the training set, for example $g=1000$ for the single-component BEC ground state, the predicted state can not be regarded as a valid physics state. When inputting a value near the training set such as $g=550$, although the shape of the predicted wave-function is not accurate, the relative error of the predicted energy is in the magnitude of $10^{-2}$. Although the effectiveness of our neural network depends on the training set, the neural network can be a fast BEC ground states generator. After training, the neural network predicts ground states much faster than imaginary evolutions. Especially for two-dimensional cases, predicting a two-component BEC using neural network takes less than a millisecond while the imaginary evolution for 8000 iterations takes about 6 seconds on the same Graphics Processing Unit (GPU).

The effectiveness of convolutional neural network for describing continuous quantum system raises some open questions. Since the ground states can be ``learned" and generated by deep convolutional neural networks, can we solve GPE without training, just having the knowledge that the ground state can be represented by the neural networks?

\begin{acknowledgments}
This work is supported by the National Natural Science Foundation of China (No.11674306, No.11774093 and No.61590932), National key R\&D program (No.2016YFA0301300 and No.2016YFA0301700). We thank the Supercomputing Center of University of Science and Technology of China for the GPU resources.  

\end{acknowledgments}

\end{document}